\documentclass[prd,superscriptaddress,amsfonts,amssymb,amsmath,showpacs]{revtex4-2}
\usepackage{bm}
\usepackage{amsfonts}
\usepackage{latexsym}
\usepackage[utf8]{inputenc}
\usepackage{graphicx}
\usepackage{amsmath}
\usepackage{palatino}
\usepackage{mathpazo}
\usepackage[british]{babel}
\usepackage{hhline}
\usepackage{multirow}
\usepackage{textcomp}
\usepackage{float}
\usepackage{booktabs}
\usepackage{dcolumn}
\usepackage{hhline}
\usepackage{multirow}
\usepackage{ragged2e}
\usepackage{hyperref}
\hypersetup{colorlinks,citecolor=blue}
\hypersetup{colorlinks=true,linkcolor=red,filecolor=magenta,    urlcolor=cyan}
\usepackage{amsmath}
\usepackage{xcolor}
\usepackage{orcidlink}
\usepackage{epsfig}
\usepackage{caption}
\usepackage{subcaption}
\usepackage{commath}
\def\jnl@style{\it}
\def\aaref@jnl#1{{\jnl@style#1}}

\def\aaref@jnl#1{{\jnl@style#1}}

\def\aj{\aaref@jnl{AJ}}                   
\def\apj{\aaref@jnl{ApJ}}                 
\def\apjl{\aaref@jnl{ApJ}}                
\def\apjs{\aaref@jnl{ApJS}}               
\def\apss{\aaref@jnl{Ap\&SS}}             
\def\aap{\aaref@jnl{A\&A}}                
\def\aapr{\aaref@jnl{A\&A~Rev.}}          
\def\aaps{\aaref@jnl{A\&AS}}              
\def\mnras{\aaref@jnl{Mon.~Not.~Roy.~Astron.~Soc.}}             
\def\prd{\aaref@jnl{Phys.~Rev.~D}}        
\def\prc{\aaref@jnl{Phys.~Rev.~C}}  
\def\prl{\aaref@jnl{Phys.~Rev.~Lett.}}    
\def\qjras{\aaref@jnl{QJRAS}}             
\def\skytel{\aaref@jnl{S\&T}}             
\def\ssr{\aaref@jnl{Space~Sci.~Rev.}}     
\def\zap{\aaref@jnl{ZAp}}                 
\def\nat{\aaref@jnl{Nature}}              
\def\aplett{\aaref@jnl{Astrophys.~Lett.}} 
\def\apspr{\aaref@jnl{Astrophys.~Space~Phys.~Res.}} 
\def\physrep{\aaref@jnl{Phys.~Rep.}}      
\def\physscr{\aaref@jnl{Phys.~Scr}}       
\def\commat{\aaref@jnl{Comm.~Math.~Phys.}}              
\def\science{\aaref@jnl{Science}}               
\def\cqg{\aaref@jnl{Classical Quant.~Grav.}}            
\def\jpcs{\aaref@jnl{JPCS}}                                     
\def\ijmpd{\aaref@jnl{Int.~J.~Mod.~Phys.~D}}                    
\def\grg{\aaref@jnl{Gen.~Relat.~Gravit.}}               
\def\rpp{\aaref@jnl{Rep.~Prog.~Phys.}}          
\def\npa{\aaref@jnl{Nucl.~Phys.~A}}        
\def\lrr{\aaref@jnl{Living Rev.~Rel.}}                   
\def\jcap{\aaref@jnl{J.~Cosmology Astropart.~Phys.}}    
\def\rmp{\aaref@jnl{Rev.~Mod.~Phys.}}   
\def\epjc{\aaref@jnl{Eur.~Phys.~J.~C}} 
\def\plb{\aaref@jnl{~Phy.~Lett.~B}} 
\def\mpla{\aaref@jnl{Mod.~Phy.~Lett.~A}} 
\def\arxiv{\aaref@jnl{arxiv.org}}

\allowdisplaybreaks[1]

\begin{document}

\title{Noether Symmetries in \texorpdfstring{$f(T,T_G)$}{} Cosmology}

\author{S. A. Kadam\orcidlink{0000-0002-2799-7870}}
\email{k.siddheshwar47@gmail.com}
\affiliation{Department of Mathematics, Birla Institute of Technology and Science-Pilani, Hyderabad Campus, Hyderabad-500078, India}

\author{B. Mishra\orcidlink{0000-0001-5527-3565}}
\email{bivu@hyderabad.bits-pilani.ac.in}
\affiliation{Department of Mathematics, Birla Institute of Technology and Science-Pilani, Hyderabad Campus, Hyderabad-500078, India}

\author{Jackson Levi Said\orcidlink{0000-0002-7835-4365}}
\email{jackson.said@um.edu.mt}
\affiliation{ Institute of Space Sciences and Astronomy, University of Malta Msida MSD 2080, Malta}
\affiliation{ Department of Physics, University of Malta, Msida MSD 2080, Malta}

\date{\today}

\begin{abstract}
All degrees of freedom related to the torsion scalar can be explored by analysing, the $f(T,T_G)$ gravity formalism where, $T$ is a torsion scalar and $T_G$ is the teleparallel counterpart of the Gauss-Bonnet topological invariant term. The well-known Noether symmetry approach is a useful tool for selecting models that are motivated at a fundamental level and determining the exact solution to a given Lagrangian, hence we explore Noether symmetry approach in $f(T,T_G)$ gravity formalism with three different forms of $f(T,T_G)$ and study how to establish nontrivial Noether vector form for each one of them. We extend the analysis made in \cite{Capozziello:2016eaz} for the form $f(T,T_{G})=b_{0}T_{G}^{k}+t_{0}T^{m}$ and discussed the symmetry for this model with linear  teleparallel equivalent of the Gauss-Bonnet term, followed by the study of two models containing exponential form of the teleparallel equivalent of the Gauss-Bonnet term. We have shown that all three cases will allow us to obtain non-trivial Noether vector which will play an important role to obtain the exact solutions for the cosmological equations.
\end{abstract}

\maketitle

\section{Introduction}

The success of general relativity (GR) in describing the standard cosmological model ($\Lambda$CDM) evolutionary processes of the Universe \cite{misner1973gravitation,Clifton:2011jh,Aghanim:2018eyx} has been overwhelming both observationally and fundamentally in many aspects. In this setting, an initial big bang scenario is driven by inflation through the dynamics of the early Universe that then leads eventually to the current state of accelerated expansion which is caused by dark energy \cite{Riess:1998cb,Perlmutter:1998np}. In $\Lambda$CDM, the dynamics of dark energy are associated with the cosmological constant $\Lambda$ which despite observational successes has had some fundamental issues within the model \cite{RevModPhys.61.1,Appleby:2018yci,Ishak:2018his}. Another core ingredient in the concordance model is cold dark matter (CDM) which acts on the galactic scale and to form the large scale structure of the Universe. However, all searches for a particle proposals for CDM remain unobserved despite many efforts \cite{Baudis:2016qwx,Bertone:2004pz}. Most recently, the $\Lambda$CDM model has been met with a new challenge from the observational sector. The Hubble tension in $\Lambda$CDM \cite{Bernal:2016gxb,DiValentino:2020zio,DiValentino:2021izs} pertains to the growing disagreement between local measurements of the Hubble constant $H_0$ (which are independent of cosmology) \cite{Riess:2019cxk,Wong:2019kwg} and predictions of the $H_0$ using early Universe data in tandem with this model \cite{DES:2017txv,Aghanim:2018eyx}. The Hubble tension continues to be reported as an increasing problem in $\Lambda$CDM \cite{Riess:2021jrx,Brout:2021mpj,Scolnic:2021amr}, and appears together with other tensions in cosmology \cite{Abdalla:2022yfr,DiValentino:2020vvd}.

The solution may be to consider further modifications to the matter sector to alter the expansion history of the Universe at different epochs. However, another way to approach this problem may be to consider alternative descriptions of gravity with the standard cosmological model \cite{Clifton:2011jh,Capozziello:2011et,CANTATA:2021ktz,Nojiri:2010wj}. By and large, these are built on corrections to the Einstein-Hilbert action \cite{Faraoni:2008mf,Capozziello:2011et}. In these settings, the gravitational field continues to interact with matter based on the curvature associated with the Levi-Civita connection, which is the source of curvature in GR \cite{misner1973gravitation,nakahara2003geometry}. On the other hand, a growing body of work considers other connections by which to express the interactions of gravity such as the torsion associated with teleparallel gravity (TG) \cite{Aldrovandi:2013wha,Cai:2015emx,Krssak:2018ywd,bahamonde2021teleparallel} through the teleparallel connection. This connection is curvature-less meaning that all measures of curvature identically vanish, such as the Ricci scalar $R=0$. This does not mean that the regular Ricci scalar vanishes ($\mathring{R} \neq 0$ - We use over-circles represent quantities calculated with the Levi-Civita connection). Relating both forms of gravity, TG can produce a torsion scalar $T$ which is equivalent to the Ricci scalar (up to a boundary term). The action based on the torsion scalar is called the \textit{Teleparallel equivalent of General Relativity} (TEGR) because it produces classical equations of motion that are dynamically equivalent to GR. The difference between GR and TEGR is that the total divergence term assimilates the quantities that would produce higher order terms if they appeared in a non-linear form.

In curvature-based theories of gravity GR has been modified in numerous directions, the most direct being that of $f(\mathring{R})$ gravity \cite{Sotiriou:2008rp,Faraoni:2008mf,Capozziello:2011et}. Similarly, TEGR can be directly generalized to $f(T)$ gravity \cite{Ferraro:2006jd,Ferraro:2008ey,Bengochea:2008gz,Linder:2010py,Chen:2010va,Bahamonde:2019zea,Duchaniya:2022rqu,Franco:2020lxx}. $f(T)$ gravity turns out to be a second order theory, unlike $f(\mathring{R})$ gravity, since the total divergence term contains the elements that would otherwise produce higher order terms in the equations of motion \cite{Cai:2015emx,Farrugia:2016qqe,Finch:2018gkh,Farrugia:2016xcw,Iorio:2012cm,Deng:2018ncg}. Another interesting scalar that emerges in TG is that of the Gauss-Bonnet scalar invariant $T_G$ \cite{Kofinas:2014owa}. As in curvature based extension to GR using the Gauss-Bonnet scalar, the TEGR action can also be modified to take an arbitrary form of this invariant through $f(T,T_G)$ gravity \cite{Bahamonde:2016kba,delaCruz-Dombriz:2017lvj,delaCruz-Dombriz:2018nvt}. In Ref.~\cite{Kofinas:2014owa} the scalar was first defined and the $f(T,T_G)$ gravity field equations presented. Then, in Ref.~\cite{Kofinas:2014aka} the Friedmann equations were derived and analyzed for certain models through a dynamical systems approach, which was complemented by works in the reconstruction section \cite{delaCruz-Dombriz:2017lvj,delaCruz-Dombriz:2018nvt} that have provided further models of interest for cosmology. In Ref.~\cite{Bahamonde:2016kba} was further studied in the cosmological context but also in other astrophysical settings which are of great interest.

In this work, we explore $f(T,T_G)$ gravity through the prism of Noether symmetries \cite{Capozziello:1996bi} which is a useful tool to analyse potentially interesting models of gravity by looking for particular symmetries in the theory. This is a core tool in probing the landscape of potentially physical models in classes of gravitational theories as general as $f(T,T_G)$ gravity since its symmetries reduce the complexity of the model space. The approach takes the point-like Lagrangian together with a symmetry of the system, which are then used in tandem with Noether's theorem to reduce the system complexity so that solutions can be found for the unknown Lagrangian functional \cite{Dialektopoulos:2019mtr}. These symmetries are intrinsically connected with conserved quantities in the model system under investigation. This approach has been used in several important settings \cite{Dialektopoulos:2018qoe,Anguelova:2018vyr} such as $f(\mathring{R})$ gravity  \cite{Capozziello:2008ch,Paliathanasis:2011jq}, scalar-tensor theories \cite{Dimakis:2017zdu,Dimakis:2017kwx,Giacomini:2017yuk,Paliathanasis:2014rja,Capozziello:2018gms}, as well as non-local theories \cite{Bahamonde:2017sdo}. In TG, the method has been used to probe $f(T)$ gravity \cite{Basilakos:2013rua}, $f(T,B)$ gravity \cite{Bahamonde:2016grb}, $f(T,T_G)$ gravity \cite{Capozziello:2016eaz}, and scalar-tensor classes of TG \cite{Dialektopoulos:2021ryi,Kadam:2022lgq,duchaniya2023dynamical,duchaniya2023Noether}. In this work, we extend the approach to the Gauss-Bonnet extension of TG. Gauss-Bonnet extensions to TEGR have been shown to produce interesting cosmologies \cite{Kofinas:2014daa,Kofinas:2014aka,Kofinas:2014owa}, as well as in the astrophysics \cite{Bahamonde:2016kba,Bahamonde:2022chq} and gravitational wave sectors \cite{Farrugia:2018gyz}. Noether symmetries offer the possibility of producing specific models motivated by conservation laws within a general class of models such as $f(T,T_G)$ gravity. For these reasons, we study this extension of TEGR to ascertain models inspired in this way.

We first open with a summary introduction to TG in Sec.~\ref{sec:intro_tg} where we review the basic properties leading up to $f(T,T_G)$ gravity. Then, in Sec.~\ref{sec:Noether symmetries} we introduce the Noether symmetry approach and its application in $f(T,T_G)$ cosmology. In Sec.~\ref{sec:Noe_symm_class}, the resulting equations of motion are used to solve for the arbitrary Lagrangian functional parameters using general ansatz models. Finally, we summarize our results in Sec.~\ref{sec:conclusion} where we describe the results in the context of the literature.

\section{Teleparallel Gravity} \label{sec:intro_tg}

The curvature of GR is replaced with torsion in TG through the use of the teleparallel rather than the Levi-Civita connection \cite{bahamonde2021teleparallel}. In this context, curvature is a product of the Levi-Civita connection $\mathring{\Gamma}^{\sigma}_{\ \ \mu\nu}$ (over-circles are used throughout to denote quantities determined using the Levi-Civita connection) rather than the metric $g_{\mu\nu}$, which is now interchanged with the teleparallel connection $\Gamma^{\sigma}_{\ \ \mu\nu}$ \cite{Hayashi:1979qx,Aldrovandi:2013wha}. The teleparallel connection continues to satisfy metricity, however, when replaced into the Riemann tensor this vanishes ($R^{\beta}_{\ \ \mu\nu\alpha} = 0$) but this is not to say that the regular Levi-Civita Riemann tensor vanishes ($ \mathring{R}^{\beta}_{\ \ \mu\nu\alpha} \neq 0$) (see reviews in Refs. \cite{Krssak:2018ywd,Cai:2015emx,Aldrovandi:2013wha}). This points to the need for different tensorial quantities to measure the effect of gravitation.

Torsion in the TG setting is more conveniently expressed through tetrads $e^{A}_{\mu}$ (and its inverses $ E_{A}^{\mu}$) which then act as agents to produce the metric through 
\begin{align}\label{metric_tetrad_rel}
    g_{\mu\nu}=e^{A}_{\mu} e^{B}_{\nu}\eta_{AB}\,,& &\eta_{AB} =  E_{A}^{\ \ \mu} E_{B}^{\ \ \nu}g_{\mu\nu}\,,
\end{align}
where Latin indices represent coordinates on the tangent space while Greek indices continue to represent indices on the general manifold \cite{Cai:2015emx}. In curvature-based gravity, tetrads also appear but they are suppressed in most circumstances \cite{Chandrasekhar:1984siy}. This may be because they are not flat in these settings, while in TG they are introduced as flat connections. Another important property of these tetrads is that they satisfy orthogonality conditions
\begin{align}
    e^{A}_{\ \ \mu} E_{B}^{\ \ \mu}=\delta^A_B\,,&  & e^{A}_{\ \ \mu} E_{A}^{\ \ \nu}=\delta^{\nu}_{\mu}\,.
\end{align}

The teleparallel connection can then be defined through\cite{Weitzenbock1923,Krssak:2015oua}
\begin{equation}
     \Gamma^{\sigma}_{\ \ \nu\mu} :=  E_{A}^{\ \ \sigma}\left(\partial_{\mu} e^{A}_{\ \ \nu} +  \omega^{A}_{\ \ B\mu} e^{B}_{\ \ \nu}\right)\,,
\end{equation}
where the tetrad embodies the degrees of freedom of the metric while the spin connection, $ \omega^{A}_{\ \ B\mu}$, takes on the six degrees of freedom of the local Lorentz invariance. Thus, the tetrad-spin connection pair produce equations of motion that embody both the Lorentz and gravitational degrees of freedom. Given that the teleparallel Riemann tensor vanishes, we define a torsion tensor through \cite{Hayashi:1979qx}
\begin{equation}
    T^{\sigma}_{\ \ \mu\nu} :=2 \Gamma^{\sigma}_{\ \ [\nu\mu]}\,,
\end{equation}
where square brackets denote the anti-symmetry operator, and where, the torsion is the measure of the anti-symmetry of the connection \cite{Aldrovandi:2013wha}. This is invariant under both local Lorentz and diffeomorphic transformations. We can also define a contortion tensor as
\begin{equation}
   K^{\sigma}_{\ \ \mu\nu} := \Gamma^{\sigma}_{\mu\nu} - \mathring{\Gamma}^{\sigma}_{\mu\nu} =\frac{1}{2}\left(T_{\mu \ \ \nu}^{\ \ \sigma}+ T_{\nu\ \ \mu}^{\ \ \sigma} - T^{\sigma}_{\ \ \mu\nu}\right)\,,
\end{equation}
which plays an important role is defining certain scalars as well as in relating curvature and torsional quantities since it is directly related to the Levi-Civita connection. Using the torsion tensor, a torsion scalar can be defined as \cite{Krssak:2018ywd,Cai:2015emx,Aldrovandi:2013wha,bahamonde2021teleparallel}
\begin{equation}\label{eq:torsion_scalar_def}
    T:=\frac{1}{4}T^{\alpha}_{\ \ \mu\nu}T_{\alpha}^{\ \ \mu\nu} + \frac{1}{2}T^{\alpha}_{\ \ \mu\nu}T^{\nu\mu}_{\ \ \  \  \alpha} - T^{\alpha}_{\ \ \mu\alpha}T^{\beta\mu}_{\ \ \ \ \beta}\,,
\end{equation}
by demanding that the action based on the torsion scalar produces the same equations of motion as the Einstein-Hilbert action (up to a total divergence term).

We again emphasize that the teleparallel connection is curvature-less and the Ricci scalar vanishes when calculated with the teleparallel connection, i.e. $R\equiv 0$. In this setting, we can thus write the regular Ricci scalar, $\mathring{R}=\mathring{R}(\mathring{\Gamma}^{\sigma}_{\ \ \mu\nu})$, as \cite{Bahamonde:2015zma,Farrugia:2016qqe}
\begin{equation}\label{LC_TG_conn}
    R=\mathring{R} + T - B = 0\,.
\end{equation}
where $B$ represents a total divergence term and is defined as
\begin{equation}\label{eq:boundary_term_def}
    B = \frac{2}{e}\partial_{\rho}\left(e T^{\mu \ \ \rho}_{\ \ \mu}\right)\,,
\end{equation}
where $e=\det\left(e^{a}_{\ \ \mu}\right)=\sqrt{-g}$ is the determinant of the tetrad. The expression in Eq.~(\ref{LC_TG_conn}) guarantees that GR and TEGR produce identical equations of motion in the classical regime, thus making them dynamically equivalent. Another interesting scalar invariant is the Gauss-Bonnet term \cite{Kofinas:2014daa,Kofinas:2014owa,Zubair:2015yma,delaCruz-Dombriz:2017lvj,delaCruz-Dombriz:2018nvt,LOHAKARE2023101164}
\begin{equation}
    G = \mathring{R}^{2} - 4\mathring{R}_{\mu\nu}\mathring{R}^{\mu\nu} + \mathring{R}_{\mu\nu\alpha\beta}\mathring{R}^{\mu\nu\alpha\beta}\,,
\end{equation}
which has been derived in the TG setting to be defined as
\begin{equation}\label{eq:T_G_def}
    T_G = \Big(K_{a \ \ e}^{\ \ i} K_{b}^{\ \ ej}K_{c \ \ f}^{\ \ k} K_{d}^{\ \ fl} - 2K_{a}^{\ \ ij} K_{b \ \ e}^{\ \ k} K_{c \ \ f}^{\ \ e} K_{d}^{\ \ fl} + 2K_{a}^{\ \ ij}K_{b \ \ e}^{\ \ k}K_{f}^{\ \ el}K_{d \ \ c}^{\ \ f} + 2K_{a}^{\ \ ij}K_{b \ \ e}^{\ \ k} K_{c,d}^{\ \ \ \ el} \Big)\delta_{iklk}^{abcd}\,,
\end{equation}
where $\delta_{iklk}^{abcd} = \epsilon^{abcd}\epsilon_{ijkl}$ is the generalized Kronecker delta function \cite{Bahamonde:2016kba}, and which also has an equivalency with the regular Gauss-Bonnet term up to a total divergence term defined as
\begin{equation}
    B_G = \frac{1}{e}\delta^{abcd}_{ijkl}\partial_{a} \left[K_{b}^{\ \ ij}\left(K_{c\ \ ,d}^{\ \ kl} + K_{d \ \ c}^{\ \ m}K_{m}^{\ \ kl}\right)\right]\,.
\end{equation}
Together, these two quantities produce the teleparallel equivalent of the Gauss-Bonnet term
\begin{equation}
    G = -T_G + B_G\,.
\end{equation}

Analogous to modifications of GR, we can consider a generalization of the Einstein-Hilbert action by adding generalizations both in terms of the torsion scalar, but also in terms of the Gauss-Bonnet term. The Gauss-Bonnet term has been related both to drivers of inflation as well as aspects of dynamical dark energy. For these reasons we consider the action \cite{Kofinas:2014daa,Kofinas:2014owa}
\begin{equation}\label{f_T_G_ext_Lagran}
    \mathcal{S}_{f(T,T_G)}^{} =  \frac{1}{2\kappa^2}\int \mathrm{d}^4 x\; e\,f(T,T_G) + \int \mathrm{d}^4 x\; e\mathcal{L}_{\text{m}}\,,
\end{equation}
where $\kappa^2=8\pi G$, and $\mathcal{L}_{\text{m}}$ is the matter Lagrangian in the Jordan frame. TEGR is recovered for the limit where $f(T,T_G) \rightarrow -T$. In this formalism, the second and fourth order contributions to the field equations are associated with the torsion scalar and Gauss-Bonnet term contributions respectively.

For a flat, homogeneous and isotropic cosmological background, the tetrad can be written as \cite{bahamonde2021teleparallel}
\begin{equation}\label{flrw_tetrad}
    e^{A}_{\ \ \mu}=\textrm{diag}(1,a(t),a(t),a(t))\,,
\end{equation}
where $a(t)$ is the scale factor, and which produces the spatially flat homogeneous and isotropic metric
\begin{equation}
    ds^2 = -dt^2+a(t)^2(dx^2+dy^2+dz^2)\,,
\end{equation}
through Eq.~(\ref{metric_tetrad_rel}). The tetrad in Eq.~(\ref{flrw_tetrad}) is compatible with the so-called Weitzenb\"{o}ck gauge where $\omega^{A}_{\ \ B\mu}=0$ \cite{Krssak:2015oua,Tamanini:2012hg}. Using the definitions in Eqs.~(\ref{eq:torsion_scalar_def},\ref{eq:T_G_def})
\begin{equation}\label{T,T_G}
    T =6H^2\,,\quad T_G = 24H^2\left(\dot{H}+H^2\right)\,,
\end{equation}
where the Gauss-Bonnet term turns out to have the same value for this background as its curvature analogue. The Friedmann equations for this set up are given by \cite{Kofinas:2014daa,Bahamonde:2016kba}
\begin{align}
    f - 12 H^2 f_T - T_G f_{T_G} + 24 H^3 \dot{f}_{T_G} &= 2 \kappa^2 \rho\\
    f - 4 (\dot{H} +3 H^2) f_{T} - 4 H \dot{f}_{T}-T_{G} F_{T_{G}} + \frac{2}{3 H} T_G \dot{f}_{T_G} + 8 H^2 \ddot{f}_{T_{G}} &= -2\kappa^2 p
\end{align}
where overdots refer to derivatives with respect to cosmic time $t$, and where $\rho$ and $p$ respectively represent the energy density and pressure of matter.

\section{Noether Symmetry Approach in \texorpdfstring{$f(T,T_G)$}{} Cosmology} \label{sec:Noether symmetries}

In this section, we explore the Noether symmetries through the point-like Lagrangian. This produces a set of equations of motion that can be solved for particular model settings. We start with the Euler-Lagrange equations which are given by
\begin{align}\label{Euler Lagrange Equations}
    \frac{d}{dt} \frac{\partial \mathcal{L}}{\partial \dot{a}}=\frac{\partial \mathcal{L}}{\partial a}\,, \quad  \quad \frac{d}{dt} \frac{\partial \mathcal{L}}{\partial \dot{T}}=\frac{\partial \mathcal{L}}{\partial T} \,, \quad  \quad \frac{d}{dt} \frac{\partial \mathcal{L}}{\partial \dot{T}_{G}}=\frac{\partial \mathcal{L}}{\partial T_{T_{G}}}\,,
\end{align}
where the energy condition can be written as
\begin{align}\label{energy condition}
    E_{\mathcal{L}}=  \frac{\partial\mathcal{L}}{\partial \dot{a}} \dot{a} + \frac{\partial\mathcal{L}}{\partial \dot{T}} \dot{T} + \frac{\partial\mathcal{L}}{\partial \dot{T_{G}}} \dot{T_{G}} - \mathcal{L}  \textbf{=0}\,,
\end{align}
which together form the dynamical system from the Lagrangian.

Let us consider the canonical variables $a$, $T$ and $T_{G}$ in order to derive the \texorpdfstring{$f(T,T_G)$}{} action through
\begin{equation}\label{Action with point like Lagrangian}
    \mathcal{S}_{f(T,T_G)}= \int \mathcal{L}(a,\dot{a},T,\dot{T},T_{G},\dot{T_{G}}) \, dt\,,
\end{equation}
where \texorpdfstring{$\mathcal{L}(a,\dot{a},T,\dot{T},T_{G},\dot{T_{G}})$}{} is the point-like canonical Lagrangian. By using Eq.(\ref{T,T_G}), we can rewrite the action equation in Eq.~(\ref{f_T_G_ext_Lagran}) into its point-like representation using the Lagrange multipliers $\lambda_{1}$ and $\lambda_{2}$ as follow \cite{Capozziello:2016eaz},
\begin{equation}\label{Action with lambda1 and lambda2}
    \mathcal{S}_{f(T,T_G)}=2 \pi^2  \int dt \bigg\{f(T,T_G) a^3 - \lambda_{1} \left[T-6\left(\frac{\dot{a}}{a}\right)^2\right]-\lambda_2\left(T_G-24\left[\frac{\dot{a}^2 \ddot{a}}{a^3}\right]\right) \bigg\}\,.
\end{equation}
Now, by varying this action with respect to $T$ and $T_G$, we find 
\begin{align}
    \left(a^3 \, f_{T}-\lambda_{1}\right)  =0 &\implies \lambda_{1}=a^3 \, f_{T}\,,\\
    \left(a^3 \, f_{T_G}-\lambda_{2}\right)  =0 &\implies \lambda_{2}=a^3 \, f_{T_G}\,.
\end{align}
Thus, the action in Eq.~(\ref{Action with lambda1 and lambda2}) can be written in following form
\begin{equation}\label{Action with values of lambda1 and lambda2}
    \mathcal{S}_{f(T,T_G)} = 2 \pi^2  \int dt \bigg\{f(T,T_G) a^3 - a^3 \, f_{T} \left[T-6\left(\frac{\dot{a}}{a}\right)^2\right]-a^3 \, f_{T_G}\left(T_G-24\left[\frac{\dot{a}^2 \ddot{a}}{a^3 }\right]\right) \bigg\}\,,
\end{equation}
\textbf{thus,} the point-like Lagrangian is given by
\begin{align}\label{point-like Lagrangian}
    \mathcal{L}_{f(T,T_G)}=a^3\left[f(T,T_{G})-T\, f_T- T_G\, f_{T_G}\right] + \, 6 a \dot{a}^2 \, f_{T} -8 \dot{a}^3 \, \dot{f}_{T_G} \,.
\end{align}

On the other hand, the Euler-Lagrange equations for the variables $a$, $T$ and $T_{G}$ can be respectively obtained to give the following \cite{Capozziello:2016eaz},
\begin{align}\label{Friedmann Eqs by Point Like Lagrangian}
    6\dot{a}^2 \, f_{T} +12 a \ddot{a} \, f_{T} + 12 a \dot{a} \, \dot{f}_T -24 \left(2 \dot{a} \ddot{a}\, \dot{f}_{T_{G}}+\dot{a}^2 \ddot{f}_{T_{G}} \right) -3a^2   f +3 a^2 T f_{T} +3 a^2 T_{G}  f_{{T}_{G}} &= 0\,,\\
    \left(-24\dot{a}^2\ddot{a}+a^3  T_{G}\right) f_{T_{G}  T}+\left( a^3 T - 6 a \dot{a}^2\right) f_{T T} &= 0\,,\\
    \left(-24\dot{a}^2\ddot{a}+a^3 \, T_{G}\right) f_{T_{G}\, T_{G}}+\left( a^3 T - 6 a \dot{a}^2\right) f_{T T_{G}} &= 0\,.
\end{align}
Now, the energy condition described in Eq.~(\ref{energy condition}) can be associated with the Lagrangian described in Eq.~(\ref{point-like Lagrangian}) to give
\begin{align}\label{Energy density Eq by Point Like Lagrangian}
    \left(6 a \dot{a}  f_{T} -24 \dot{a} ^2 \dot{f}_{T_{G}}\right) \dot{a}-a^3 f + a^3 T_{G} f_{T_{G}}+ a^3 T f_{T} = 0\,.
\end{align}
For the configuration space $\mathcal{Q}$=$(a,T,T_{G})$ of the point-like Lagrangian in Eq.(\ref{point-like Lagrangian}), suppose that the dynamics of a system are governed by a Lagrangian $\mathcal{L}$ in terms of $n$ generalized coordinates $q^{i}$ with $t$ as an independent variable. The infinitesimal point transformations operated on the system can be expressed as follow, \\
\begin{align}
    t \implies t'=t+\epsilon \xi (q^{i},t),   \quad q^{i}\implies q^{i'}=q^{i}+\epsilon \eta^{i}(q^{i},t) \,,
\end{align}
the generating vector in this case can be written as,\\
\begin{align}\label{generating vector}
    \chi=\xi (t,a,T,T_{G}) \partial_{t}+\sum \eta_{q^i} (t,a,T,T_{G}) \partial_{q^i}      \quad q^i=a,T,T_{G} \,.
\end{align}
For any differentiable function $F$, the action of this transformation is given by,\\
\begin{align}
F(q',t')=F(q,t)+\epsilon \chi(F(q,t))+\mathcal{O}(\epsilon^{2})\,.
\end{align}
this can be extended readily for the case where $F$ shows velocity dependence as follow,\\
\begin{align}
F(q',\dot{q}',t')=F(q,\dot{q},t)+\epsilon \chi^{(1)} (F(q,\dot{q},t))+\mathcal{O}(\epsilon^{2})\,,
\end{align}
for the system with Lagrangian $\mathcal{L}=\mathcal{L}(t,q,\dot{q})$, the action is said to invariant under infinitesimal transformations if the Rund-Trautman identity holds \cite{Basilakos:2011rx}, 

\begin{align}\label{Rund-Trautman identity}
    \chi^{(1)} \mathcal{L}+\frac{d\xi(t,q^i)}{dt} \mathcal{L}=\frac{dg(t,q^i)}{dt}\,.
\end{align}

Where,
\begin{align}
\chi^{(1)}=\chi+(\dot{\eta}^i-\dot{q}^i \dot{\xi}) \frac{\partial}{\partial \dot{q}^i}\,,
\end{align}
the second prolongation vector can be written as,\\
\begin{align}
\chi^{(2)}=\chi^{(1)}+(\ddot{\eta}^i-\dot{q}^i \ddot{\xi}-2\ddot{q}^{i}\dot{\xi}) \frac{\partial}{\partial \dot{q}^i} \,.   
\end{align}
We can generalise this up to the $nth$ prolongation $\chi^{(n)}$ together with the independent variable cosmic time $t$.

\section{Study of forms of \texorpdfstring{$f(T,T_G)$}{} Gravity using Noether symmetries}\label{sec:Noe_symm_class}

Here, we apply the Rund-Trautman identity, which is presented in Eq.~(\ref{Rund-Trautman identity}), to the point-like Lagrangian in Eq.~(\ref{point-like Lagrangian}). By comparing the coefficients of the terms $a$, $\dot{a}$, $T$, $\dot{T}$, $T_G$, $\dot{T_G}$,  $\dot{a}\dot{T}$, $\dot{a}\dot{T_G}$, and so on, we obtain a system of 21-differential equations, which are
\begin{align}\label{first diff eq}
    \eta_{a} 6 f_{T}+\eta_{T} 6 a f_{T,T} + \eta_{T_G} 6 a f_{T,T_{G}} -6 a f_{T} \xi_{,t} +12 a f_{T} \eta_{a,a}&=0\\
  -\eta_{T} 8 f_{T_G, T, T_G}-\eta_{T_G} 8 f_{T_G,T_G,T_G}-24 \eta_{a,a} f_{T_G,T_G} +24 \xi_{,t}f_{T_G,T_G}-8 f_{T_G,T} \eta_{T,T_G} -8 f_{T_G,T_G}\eta_{T_G,T_G}&=0 \label{Constraint Eq on k}\\
   - \eta_{T} 8 f_{T_G,T,T}-\eta_{T_G} 8 f_{T_G,T_G,T}-24 \eta_{a,a} f_{T_G,T}+24\xi_{,t}f_{T_G,T}-8f_{T_G,T} \eta_{T,T}-8f_{T_G,T_G} \eta_{T_G,T}& =0\\
 -8f_{T_G,T} \eta_{T,t}-8\eta_{T_G,t}f_{T_G,T_G}-6af_{T}\xi_{,a}&=0\\
    -12 a \xi_{,T}-24\eta_{a,t} f_{T_G,T}+6 a f_{T} \xi_{,T}&=0\\
   -8f_{T_G,T}\eta_{T,a}-8f_{T_G,T_G} \eta_{T_G,a}&=0\\
   24\xi_{,T} f_{T_G,T_G}+24\xi_{,T_G}f_{T_G,T}&=0\\
    -24\eta_{a,T} f_{T_G,T_G} -24\eta_{a,T_G}f_{T_G,T}&=0\\
   -6a f_{T}\xi_{,T_G} -24\eta_{a,t} f_{T_G,T_G}&=0\\
    12 a f_{T} \eta_{a,T}&=0\\
  12 a f_{T} \eta_{a,T_G}&=0\\
    -24\eta_{a,T} f_{T_G,T}&=0\\
    -24\eta_{a,T_G} f_{T_G,T_G}&=0\\
    24\xi_{,T_G} f_{T_G,T_G}&=0\\
    24\xi_{,a}f_{T_G,T}&=0\\
    24\xi_{,a} f_{T_G,T_G}&=0\\
    24\xi_{,T} f_{T_G,T}&=0\\
    \eta_{a,t} 12 a f_{T} + a^3 f \xi_{,a} - a^3 T f_{T} \xi_{,a}-a^3 T_{G} f_{T_G} \xi_{,a}&=g_{,a}\\
      a^3 f \xi_{,T} -a^3 T f_T \xi_{,T} - a^3  T_G f_{T_G} \xi_{,T} &=g_{,T}\\
    a^3 f \xi_{,a}-a^3 T f_T \xi_{,T_G}-a^3 T_G f_{T_G}\xi_{,T_G} &=g_{,{T_G}}\\
     3a^2 \eta_a f-3a^2 \eta_a T f_T-\eta_a 3a^2 T_G f_{T_G}-\eta_T a^3 T f_{TT}-\eta_{T}a^3 T_{G} f_{T_G, T} + a^3 f \xi_{,t}-a^3 T f_T \xi_{,t}-\nonumber\\ a^3 T f_{T,T_G} \eta_{T_G}-a^3 T_{G} f_{T_G}\xi_{,t}-\eta_{T_G}a^3 T_{G} f_{T_G,T_G} &=g_{,t}\label{last diff eq}
\end{align}

Now, to find the Noether symmetries, we need to find the coefficients of the generating vector described in Eq.~(\ref{generating vector}) such that symmetry condition in Eq.(\ref{Rund-Trautman identity}) satisfy. For this one can obtain the solution to the above system of equations by approaching to two ways, either by calculating coefficients of the Noether vector by considering the form of $f(T, T_{G})$ gravity or can obtain the solution in the form of $f(T, T_{G})$ function by considering the form of the Noether vector. We consider some form of $f(T,T_{G})$ in the subsequent section and obtain the  general form of coefficients of the Noether vector. We also discus the possible exact solution for each case which play an important role to enlighten the cosmological implications of Noether symmetry approach in $f(T,T_G)$ gravity formalism.

\subsection{The case \texorpdfstring{$f(T,T_{G})=b_{0}T_{G}^{k}+t_{0}T^{m}$}{}}

In this case, we substitute this form of $f(T,T_{G})$ in the system (\ref{first diff eq}-\ref{last diff eq}) for the particular case $k=1$ and arbitrary $m$ i.e. $f(T,T_{G})=b_{0}T_{G}+t_{0}T^{m}$. In this case the Noether vector coefficient $\xi$ and function $g$ remain as a constant functions denoted by $\tau_{1}$ and $\tau_{2}$ respectively, the Noether symmetry vector assumes the form as follow, 
\begin{align}\label{Noether vector M-I}
    \chi=\left(\xi=\tau_{1}, \quad \eta_{a}=\alpha_{0}a^{1-\frac{3}{2m}},  \quad  \eta_{T_G}=\eta_{T_G}(t,a,T,T_{G}),  \quad \eta_{T}=-\frac{3\alpha_{0}Ta^{-\frac{3}{2m}}}{m}, \quad g=\tau_{2}\right)\,.
\end{align}
Along with this case, the mixed power law form of teleparallel torsion scalar $T$ and the teleparallel equivalent of the Gauss-Bonnet term $T_{G}$ is also discussed in Ref.~\cite{Capozziello:2016eaz}. Let us find the cosmological solutions for the separated power law model, $f(T,T_G)=b_{0}T_{G}+t_{0}T^m$. The pointlike Lagrangian presented in Eq.~(\ref{point-like Lagrangian}) takes the form as,\\
\begin{align}\label{Lagrangian for Model I}
     \mathcal{L}=a^3(1-m) T^m t_0+6 a \dot{a}^2 t_0 m T^{m-1}\,.
\end{align}
The Euler-Lagrange equation for $T_G$ can give the Lagrange multiplier $T_G$ and the Euler-Lagrange equation for variable $a$ in Eq.~(\ref{Euler Lagrange Equations}) and the energy condition equation in Eq.(\ref{energy condition}) respectively can be written as,
\begin{align}\label{eq:61}
    12 t_0 m \left(\dot{a}^2 T^{m-1}+a \ddot{a} T^{m-1}+a \dot{a} (m-1) T^{m-2} \dot{T}\right)-3a^2 (1-m) T^m t_0 - 6\dot{a}^2 t_0 m T^{m-1} =0\,,\\
    6 a \dot{a}^2 t_0 m T^{m-1} - a^3 (1-m) T^m t_0=0\label{eq:62}\,.
\end{align}

Using Eq.~(\ref{T,T_G}), the above equations in Eqs.~(\ref{eq:61}--\ref{eq:62}) can be reformulated to take the form 
\begin{align}\label{eq:63}
    6^m (2 m-1) t_{0}  \left(\frac{\dot{a}^2}{a^2}\right)^{m-1} \left((3-2 m) \dot{a}^2+2 m a \ddot{a}\right)=0\,,\\
    6^m (2 m-1) t_{0}  a^3 \left(\frac{\dot{a}^2}{a^2}\right)^m=0\,,\label{eq:64}
\end{align}
where the scale factor features more prominently.

As discussed in Ref.~\cite{Basilakos:2013rua}, the differential equation Eq.~(\ref{eq:63}) admit the following solution
\begin{align}
    a(t)=\tilde{c}_2 (3 t-2 \tilde{c}_1 m){}^{\frac{2 m}{3}} \,, \quad \text{for} \quad m\neq\frac{1}{2}\,,
\end{align}
 where $\tilde{c}_1,\tilde{c}_2$ are the integrating constants, one can also note that both the differential equations Eqs.~(\ref{eq:63}--\ref{eq:64}) will satisfy the equality for $m=\frac{1}{2}.$ In this case, the deceleration parameter takes the form $-1+\frac{3}{2m}$. Hence this exact solution can explain the accelerating Universe for the parameter $m$, within the range $m<0 \quad \text{or} \quad m>\frac{3}{2}$. Moreover, these cosmology expressions admit de-Sitter solution as follow,
 
\begin{align}
    a(t)=  e^{s t}\,, \quad \text{where} \quad \left(s\neq 0\land \left(m=\frac{1}{2}\lor t_{0} =0\right)\right)\lor (\Re(m)>0\land s=0)\,.
\end{align}
Here $s, t_{0}$ are the real constants. In this case the solution holds conditions on the model parameter  $m, t_{0}$ and parameter $s$ such that for nonzero $s$ we get $m=\frac{1}{2}$ similar condition obtained in the power law solution and for $s=0$, the expressions will admit solution at $\Re(m)>0$,  $\Re(m)$ represents the real part of the parameter $m$.
\subsection{The case \texorpdfstring{$f(T,T_{G})=-T+je^{m T_G}$}{}}\label{Case-B}

To study the analysis of critical points representing different phases of the evolution of the Universe, the exponential form of teleparallel torsion scalar model has been studied in the dynamical system approach in Ref.~\cite{Mirza_2017}. In this case to retrieve the GR limit the model contains the linear torsion term with negative sign along with the exponential form of scalar multiple of the teleparallel equivalent of the Gauss-Bonnet term $T_G$. The coefficient $\xi$ in this case constrained in such a way that it satisfies the conditions obtained while solving system of equations from Eqs.~(\ref{first diff eq}--\ref{last diff eq}). The Noether vector coefficient, $\xi$ assumes the linear form of time $t$, with the arbitrary parameters $k_{1}$ and $k_{2}$ and the constant $c_{1}$ is the integration constant. Here, to solve above differential equations system we arrive on the conclusion that $\eta_{T_G}$ takes the specific form $\eta_{T_G}=\eta_{T_G}( T_G)$ and the coefficient $\eta_{a}, \eta_{a}=\eta_{a}(a)$. The function $g$ remains constant and represented using arbitrary constant $\tau_2$. With these conditions we have obtained the nontrivial form for Noether vector, giving
\begin{align}\label{Noether vector M-II}
    \chi=\left(\xi=k_{1}t+k_{2},\quad \eta_{a}=\frac{a k_1}{3},  \quad  \eta_{T_G}=\frac{2k_1}{m}+c_1 e^{-m T_{G}},  \quad \eta_{T}=\eta_{T}(t,a,T,T_{G}),\quad g=\tau_{2}\right)\,.
\end{align}
In this case the Lagrangian Eq.~(\ref{point-like Lagrangian}) takes the form
\begin{align}\label{Lagrangian for Model II}
     \mathcal{L}=-6
 a  \dot{a}^2+j e^{m T_G}\left(a^3-T_G a^3 m-8 \dot{a}^3\dot{T}_G m^2\right)\,,
\end{align}
where we again use Eq.~(\ref{point-like Lagrangian}).

Using the Euler-Lagrange equations in Eq.~(\ref{Euler Lagrange Equations}) for the variable $a$ and the energy condition in Eq.~(\ref{energy condition}), respectively give,
\begin{align}\label{eq:68}
    -6\dot{a}^2-12 a \ddot{a}-24j m^2 e^{m T_G}\left(m \dot{T}_G \dot{a}^2 \dot{T}_G+2\dot{a}\ddot{a}\dot{T}_G+\dot{a}^2\ddot{T}_G\right)-j e^{m T_G}\left(3a^2-3ma^2 T_G\right)=0\,,\\
    -6 a \dot{a}^2-j  e^{m T_G}\left(24 \dot{a}^3 \dot{T}_G m^2 + a^3 -T_G a^3 m\right)=0\,.\label{eq:69}
\end{align}

The cosmological solution in this case can then be obtained to give
\begin{align}
    a(t)=e^{\lambda  t} \quad \text{where} \quad 24 \lambda ^4 m-1\neq 0 \quad  \text{and} \quad j=\frac{6 \lambda ^2 e^{-24 \lambda ^4 m}}{24 \lambda ^4 m-1}\,,
\end{align}
which is an exact solution. Here $\lambda$ is the real constant.

\subsection{The case \texorpdfstring{$f(T,T_{G})=-T+je^{m \sqrt{T_G}}$}{}}\label{Case-c}

This model takes the exponential form of scalar multiple of square root of the teleparallel equivalent of the Gauss-Bonnet term along with the linear form of torsion term with negative sign and used in the study of dynamical system analysis in teleparallel gravity formalism in Ref.~\cite{Mirza_2017}. Also the study of this form of model, successfully explains the transition phase from early to late time of cosmic evolution in teleparallel gravity formalism \cite{Linder:2010py}. As we have discussed in the previous case in this case also we obtain $\eta_{T_G}$ takes the specific form $\eta_{T_G}=\eta_{T_G}( T_G)$ and the coefficient $\eta_{a}, \eta_{a}=\eta_{a}(a)$, the other coefficients of Noether vector takes the form as follow, 
\begin{align}\label{Noether vector M-III}
    \chi=\left(\xi=k_{1}t+k_{2}, \quad \eta_{a}=\frac{a k_1}{3},  \quad  \eta_{T_G}=-\frac{4 k_1 T_{G}}{1-m \sqrt{T_{G}}}+\frac{c_1 (T_{G})^{3/2} e^{-m \sqrt{T_{G}}}}{1-m \sqrt{T_{G}}},  \quad \eta_{T}=\eta_{T}(t,a,T,T_{G}),\quad g=\tau_{2}\right)\,,
\end{align}
Here $c_{1}$ is the integration constants, while $\tau_{2}$, $k_{1}$ and $k_{2}$ are the arbitrary constants. In order to compare with the above square root free exponential form of the  teleparallel equivalent of the Gauss-Bonnet term model study, we can observe that the coefficients $\xi$, $\eta_{a}$, $\eta_{T}$ and function $g$ assumes the same form, but the coefficient $\eta_{T_G}$ will give difference in the form of the Noether symmetry generating vector.
In this case the Lagrangian Eq.(\ref{point-like Lagrangian}) takes the form as follow,
\begin{align}\label{Lagrangian for Model III}
     \mathcal{L}=-6
 a  \dot{a}^2+ e^{m \sqrt{T_G}}\left(a^3 j-\frac{\sqrt{T_G} j m a^3}{2}+\frac{2\dot{a}^3 \dot{T}_{G} j m}{(T_G)^{\frac{3}{2}}}-\frac{2(\dot{a})^{3} j m^2 \dot{T}_G}{T_G}\right) \,,
\end{align}

The Euler-Lagrange equations from Eq.~(\ref{Euler Lagrange Equations}) and energy condition from Eq.~(\ref{energy condition}) respectively can be expressed as
\begin{align}\label{eq:73}
    e^{m\sqrt{T_G}}\left((T_G)^{\frac{-3}{2}} \left(\frac{3 m^2 j (\dot{T}_G)^2 \dot{a}^2}{\sqrt{T_G}}+12\dot{a} \ddot{a}\dot{T}_G j m + 6 \dot{a}^2 \ddot{T}_G j m\right) -9\dot{a}^2 (\dot{T}_G)^2 j m (T_G)^{\frac{-5}{2}}\right)(1-\sqrt{T_G} m)\nonumber\\
    +e^{m\sqrt{T_G}}\left(-\frac{3\dot{a}^2 (\dot{T}_G)^2 j m^2}{(T_G)^2}-3a^2 j +\frac{3a^2\sqrt{T_G} j m}{2}\right)-6\dot{a}^2 -12 a \ddot{a}=0\,,\\
    -6 a \dot{a}^{2}+e^{m \sqrt{T_G}}\left(\frac{6\dot{a}^3 j m \dot{T}_G}{T_G}\left(\frac{1-m \sqrt{T_G}}{\sqrt{T_G}}\right)-a^3 j + \frac{\sqrt{T_G}j m a^3}{2}\right)=0\,.\label{eq:74}
\end{align}

The cosmological solution in this case can be obtained as follow,
\begin{align}
    a(t)=e^{\zeta  t} \quad \text{where} \quad \sqrt{6}-6 \zeta ^2 m\neq 0\quad \text{and} \quad j=-\frac{6 \sqrt{6} \zeta ^2 e^{-2 \sqrt{6} \zeta ^2 m}}{\sqrt{6}-6 \zeta ^2 m}\,,
\end{align}
which is again an exact solution. In both the cases, Case-\ref{Case-B} and Case-\ref{Case-c}, we get the de-Sitter solution with different condition on the model parameters $j$ and $m$. \\

\section{Conclusion}\label{sec:conclusion}

The symmetries are not only a mathematical tool to solve the dynamics but also their existence allows for the selection of analytical models in the cosmological aspect. The torsion scalar ($T$) and the Gauss-Bonnet topological invariant term ($T_G$) both are the functions of scale factor and its derivatives of first and second order, and so, its dynamics is not a priori canonical. In order to make the point-like Lagrangian canonical, the Lagrange multiplier has a key role. This allows to deal with the teleparallel invariant such as, $T$ and $T_G$ by using the auxiliary effective scalar fields related to $f(T, T_G)$.   In this work, we explore $f(T,T_G)$ gravity through the prism of Noether symmetries \cite{Capozziello:1996bi} which is a useful tool to analyse potentially interesting models of gravity by looking for particular symmetries in the theory. This is a core tool in probing the landscape of potentially physical models in classes of gravitational theories as general as $f(T,T_G)$ gravity since its symmetries reduce the complexity of the model space. We defined the point-like Lagrangian presented in Eq.(\ref{point-like Lagrangian}) for the action equation in Eq.(\ref{f_T_G_ext_Lagran}) of $f(T,T_G)$ gravity, and confirm its validity by recreating the Friedmann equations along with the torsion scalar and the  teleparallel equivalent of the Gauss-Bonnet term presented in Eqs.(\ref{Friedmann Eqs by Point Like Lagrangian}-\ref{Energy density Eq by Point Like Lagrangian}). In this study, we analyse the Noether symmetry approach introduced in \cite{Basilakos:2011rx} and consider the Rund-Trautman identity presented in Eq.(\ref{Rund-Trautman identity}) by which we have obtained the system of differential equations presented in Sec-(\ref{sec:Noe_symm_class}).

We also study three forms of $f(T,T_G)$ in Noether symmetry approach and confirms that all three forms admits symmetry by which we can get reduction in the complexity of the dynamical system. In the first case, we have obtained the cosmological solution to show how the presence of symmetry enables for a proper reduction of the dynamical system.  We have shown that this form also admits Noether symmetry, the nontrivial Noether symmetry generating vector is presented in Eq.(\ref{Noether vector M-I}). In the second case, we have analysed the Noether symmetry approach for exponential form of the teleparallel equivalent of the Gauss-Bonnet term which was studied in teleparallel gravity formalism in Ref(\cite{Mirza_2017}). This form also admits the Noether symmetry and the nontrivial Noether symmetry generating vector can be obtained as presented in Eq.(\ref{Noether vector M-II}). In this case, the exact cosmological solution is the de-Sitter solution. In the last case, the nontrivial Noether vector can be obtained as presented in Eq.(\ref{Noether vector M-III}). This form admits de-Sitter solution with some conditions on the model parameter $m, \zeta$ which is also an exact solution.
Further other physically viable forms of $f(T,T_G)$ can be analysed in the Noether symmetry approach which may reduce the complexity of the dynamics of the system and would be helpful to study the cosmological solutions in the $f(T,T_G)$ theory. 

The models that emerge from this analysis seems to show promise in describing the critical points that are known to have taken place during the evolution of the Universe. However, more work now needs to be undertaken to understand how these critical points, and model parameter freedoms, correlate with real observational measurements. In particular, it would be interesting to probe the early Universe in these models, which may be better able to discriminate between viable parameter ranges.

\section*{Acknowledgements}
SAK acknowledges the financial support provided by University Grants Commission (UGC) through Senior Research Fellowship (UGC Ref. No.: 191620205335) to carry out the research work. BM acknowledges IUCAA, Pune, India for hospitality and support during an academic visit where a part of this work has been accomplished. JLS would like to acknowledge funding from Cosmology@MALTA which is supported by the University of Malta. The authors are thankful to the anonymous referees for their comments and suggestions to improve the quality of the paper.

\bibliographystyle{utphys}
\bibliography{references}

\end{document}